# Type Ia Supernova Scenarios and the Hubble Sequence

P. Ruiz-Lapuente[1,2], A. Burkert[3], and R. Canal[2]

*Running title:* SNe Ia and Hubble Sequence

astro-ph/9505090  19 May 95

[1]Max–Planck–Institut für Astrophysik, Karl–Schwarzschild–Strasse 1, D–85740 Garching, Federal Republic of Germany. E–mail: pilar@MPA–Garching.MPG.DE

[2]Department of Astronomy, University of Barcelona, Martí i Franqués 1, E–08028 Barcelona, Spain. E–mail: pilar@mizar.am.ub.es, ramon@farcm0.am.ub.es

[3]Max–Planck–Institut für Astronomie, Königstuhl 17, D–69117 Heidelberg, Federal Republic of Germany. E–mail: andi@mpia-hd.mpg.de

# ABSTRACT

The dependence of the Type Ia supernova (SN Ia) rate on galaxy type is examined for three currently proposed scenarios: merging of a Chandrasekhar–mass CO white dwarf (WD) with a CO WD companion, explosion of a sub–Chandrasekhar mass CO WD induced by accretion of material from a He star companion, and explosion of a sub–Chandrasekhar CO WD in a symbiotic system. The variation of the SNe Ia rate and explosion characteristics with time is derived, and its correlation with parent population age and galaxy redshift is discussed. Among current scenarios, CO + He star systems should be absent from E galaxies. Explosion of CO WDs in symbiotic systems could account for the SNe Ia rate in these galaxies. The same might be true for the CO + CO WD scenario, depending on the value of the common envelope parameter. A testable prediction of the sub–Chandrasekhar WD model is that the average brightness and kinetic energy of the SN Ia events should increase with redshift for a given Hubble type. Also for this scenario, going along the Hubble sequence from E to Sc galaxies SNe Ia events should be brighter on average and should show larger mean velocities of the ejecta. The observational correlations strongly suggest that the characteristics of the SNe Ia explosion are linked to parent population age. The scenario in which WDs with masses below the Chandrasekhar mass explode appears the most promising one to explain the observed variation of the SN Ia rate with galaxy type together with the luminosity–expansion velocity trend.

*Subject headings:* cosmology: distance scale — galaxies: elliptical and lenticular — galaxies: spiral — stars: binaries: close — stars: supernovae: general — stars: white dwarfs

## 1. Introduction

Type Ia supernovae (SN Ia) have always been regarded as a very homogeneous class as compared with the other SN types: SNe II and SNe Ib/c. Recent observations suggest, however, that there are intrinsic differences among SN Ia and, in addition, that they might be correlated with parent–galaxy type. Filippenko (1989) and Branch & van den Bergh (1993) find such a dependence for the expansion velocities of the ejecta: in early–type galaxies the velocities are lower than in late–type galaxies, the difference between the most extreme cases being $\sim 5,000$ $km$ $s^{-1}$. Phillips (1993), Maza et al. (1994), and Hamuy et al. (1995) find a significant dispersion in the absolute magnitudes of SNe Ia at maximum light. Again, a dependence on galaxy type seems to exist in luminosity (van den Bergh & Pazder 1992), the SN Ia in late–type galaxies being brighter on average by $\sim 0.3$ mag than those in early–type galaxies (Hamuy et al.1995). Besides, the rate of SN Ia per unit of K luminosity of the parent galaxy appears to be larger (by a factor $\simeq$ 5–10) in late–type galaxies than in early–type galaxies (Della Valle & Livio 1994). These authors suggest the possibility that SN Ia might come from two different classes of progenitors: SN Ia in late–type galaxies would come from cataclysmic variable (CV) systems, in which massive WDs accrete matter at high rates, whilst SN Ia in early–type galaxies would be produced by the merging of WDs in close binary systems (CBS) (Iben & Tutukov 1984) or by WDs that explode before reaching the Chandrasekhar mass (Livne & Glasner 1991; Ruiz–Lapuente et al. 1993; Woosley & Weaver 1994). Merging of WDs has recently been reexamined by Tutukov & Yungelson (1994) to study the evolution of the SN Ia rate for different parent–galaxy types. In this *Letter* we present a preliminary analysis of the different ways in which galaxy type (through metallicity and population age effects) influences the characteristics of the progenitors of the SN Ia. Differences in the progenitors provide a natural explanation for the observed dependence of SN Ia characteristics on Hubble type.

## 2. Metallicity and Population Age Effects

Early–type galaxies have, in general, higher metallicities than late–type galaxies, but by a factor of 3–5 at most (Roberts & Haynes 1994). The peak luminosities, the explosion energies, and the rate of occurrence of SN Ia depend on the parameters characterizing the binary system at birth (the mass of the primary $M_1$, the mass ratio $q \equiv M_2/M_1$ of the secondary to the primary, and the primordial semimajor axis $A_0$), on the explosion mechanism, and on thermonuclear burning life times. Of these, only the nuclear burning life time of the secondary, which determines, at least partially, the epoch of explosion in all scenarios, does somewhat depend on metallicity within the above range. That changes the absolute and relative SN Ia rates deduced for different scenarios, and therefore must be taken into account. In the standard picture of the evolution of galaxies of the different Hubble types, star formation in elliptical (E) galaxies takes place in a burst, with a typical time scale of $\sim 1$ Gyr, while in late–type galaxies it is a continuous process lasting from the epoch of galaxy formation up to present. We will adopt, for our estimates, a time interval of $\sim 10$ Gyr since the burst of star formation in E galaxies up to now. Changes in this quantity do not affect our qualitative conclusions.

There are, within the mass–accreting WD paradigm, two different broad types of scenarios: one involving a CO WD accreting matter from a nondegenerate companion (the *single degenerate* –SD– scenario), and another in which a CO WD accretes from a degenerate companion, most likely a second CO WD (the *double degenerate* –DD– scenario). There are also two main classes of models for the explosion: in one the CO WD explodes when it reaches the Chandrasekhar mass, due to C ignition at its center, while in the other a CO WD, with a mass still lower than the Chandrasekhar mass, explodes due to a C detonation induced by the previous detonation of He accumulated at the surface (either directly accreted or resulting from burning of previously accreted H).

*1. Single Degenerate (SD) Scenario: a) Symbiotic systems.* (Munari & Renzini 1992; Kenyon et al. 1993). A hot WD accretes matter from the low–velocity wind of a red giant. The clock of the system is the evolutionary life time of the secondary. We have computed the evolution of the SN Ia rate from binary systems in E galaxies with our scenario code coupled with a Monte Carlo numerical simulation. For the IMF we have adopted that of Scalo (1986), for the mass ratio ($q$) distribution the one from Duquennoy & Mayor (1991), and for the distribution of primordial semimajor axes $A_0$ that deduced from the distribution of periods in the same source. We include only systems with $1\ yr \leq P_{orb} \leq 10\ yr$, which seem to be within the appropiate range for symbiotics. The relationship between WD mass and initial primary mass is that of Iben & Renzini (1983), adopting the value 0.6 for the mass–loss parameter $\eta$. The nuclear burning life times are from Schaller et al. (1992). The result is shown in the upper panel of Figure 1. The SN Ia rate sharply rises after $\sim 10^8$ yr as soon as the most massive secondaries ($M_2 \simeq 8 M_\odot$) end their evolution, followed by a smoother increase up to a maximum which is reached $\sim 10^9$ yr after the end of star formation. The rate declines later on, jumps to a second maximum at $\sim 5 \times 10^9$ yr, and then decreases again up to present. The rate scale corresponds to an outburst with a total mass adjusted to give a SN Ia rate $\nu = 0.003\ yr^{-1}$ in the *double degenerate* scenario (see below), 10 Gyr after the end of star formation. Shown in the second panel of Figure 1 is the time evolution of the average mass of the exploding WDs: it monotonously decreases from its maximum ($\simeq 1.2 M_\odot$), when the first SNe Ia start exploding, down to $\simeq 0.7 M_\odot$ at the present time. Of course, changes in the value of $\eta$ would shift those values. In the third panel of Figure 1, the time evolution of the average velocity of the ejecta is also shown. Therefore we expect that, *for this scenario and model, average SN Ia explosions should be dimmer at the present epoch than earlier and the explosion energies should be lower. The luminosity of SNe Ia and their explosion energy should be lower in early–type galaxies*, where the star formation process stopped long ago, *than in late–type galaxies*, where stars are still forming today. From the calculations of the dependence of the $^{56}$Ni yield on WD mass by Woosley & Weaver (1994) and Ruiz–Lapuente et al. (1993), it appears that this trend could easily explain the observed magnitude differences. The ejecta of the brighter explosions would also move faster and variations in the velocities of the Si II $\lambda$6,335 line with an amplitude $\sim 5,000$ $km$ $s^{-1}$ might be accounted for by the models (Woosley & Weaver 1994).

*b) WD + He star systems.* In the systems where a CO WD is accreting mass from a nondegenerate He star, the WD is formed after a common envelope phase which takes place when the primary becomes an AGB star (Iben & Tutukov 1991). The semimajor axis of the system is reduced. A common envelope forms again when the secondary overfills its Roche lobe and exposes its nondegenerate He core. The final separation is proportional to the square of $\alpha$, the common–envelope parameter, of order unity, related to the efficiency with which orbital energy is deposited into the ejection of the common envelope. After that, the orbital separation will decrease due to the emission of gravitational wave radiation (Landau & Lifschitz 1975), which keeps the system in contact. When the evolution of the rate is computed for an starburst population as in the previous case, one obtains the behaviour displayed in the fourth panel of Figure 2: for all values of $\alpha$, the maximum rate is reached $\simeq 2 \times 10^8$ $yr$ after the end of star formation and decreases very fast afterwards. The average WD mass (not displayed in the Figure) again shrinks with time, as in the symbiotic scenario. In summary, for WD + He star systems *the SN Ia rate decreases with time* after reaching its peak very soon after the end of star formation and *the average SN Ia becomes dimmer and less energetic*. The early maximum and the fast decline afterwards, however, disfavour this scenario as a significant contributor to the current SN Ia rate in ellipticals, although it might be at work in late–type galaxies.

*2. Double Degenerate (DD) Scenario.* In the preferred DD scenario (Iben & Tutukov

1984), two CO WDs of roughly the same mass merge by emission of gravitational wave radiation. Their total mass must exceed the Chandrasekhar mass ($\simeq 1.4 M_\odot$). Two previous common-envelope episodes are again involved. For all WD masses, the $^{56}$Ni yield and the explosion energy adopted correspond to an average value, given the lack of predictions for their possible dependence on the characteristics of the system. In all cases, the explosion takes place when the more massive WD, after tidal disruption of the less massive one, accretes enough material to reach the Chandrasekhar mass. In the fifth panel of Figure 1 we show the evolution of the SN Ia rate, for this scenario, after an outburst of star formation: depending on the value of $\alpha$, the maximum is reached $\sim 10^8\ yr$ after the end of the outburst and the rate decreases rapidly afterwards (the slope of the descent being steeper for decreasing values of $\alpha$). *The SN Ia rate in E galaxies, in the DD scenario, decreases with time, the luminosity and energy of each event remaining constant* (within the precedent assumptions). We found that DD coming from "wide" primordial systems should still be merging $\sim 10^{10}$ yr after the end of star formation in E galaxies. That, however, strongly depends on the value of the common–envelope parameter $\alpha$. From the numerical simulations of common envelope evolution by Livio & Soker (1988) and Taam & Bodenheimer (1989), it would appear that $\alpha \simeq 0.3 - 0.6$, but arguments for higher values are given by Yungelson et al. (1994) and Tutukov & Yungelson (1994). All this indicates that, even if nonzero, the rate of DD merging in E galaxies should currently be low. Note that for the adopted IMF and $q$ and $A_0$ distributions, the SN Ia rate from SD (symbiotic) systems should currently be $\sim 20$ times higher, in E galaxies, than that from DD systems (for a common–envelope parameter $\alpha = 1$ and an accretion efficiency of 100% in both DD systems and symbiotics). The contrast would be 1.5 times smaller for the Salpeter (1955) IMF and the $q$ and $A_0$ distributions adopted by Tutukov & Yungelson (1994) (see Figure 2). On the other hand, the total mass involved in the star formation outburst required to reproduce the 0.003 yr$^{-1}$ rate is about 1.5 times larger with the Salpeter IMF than with the Scalo IMF. Concerning metallicity, reducing it by a factor of 20 (from Z = 0.02 to Z = 0.001), all other parameters being equal, would increase the maximum DD rate only by 11% (for $\alpha = 1$), but it would increase the symbiotic rate at 10 Gyr by a factor 30. Concerning the absolute scale in SNe Ia rates in Figures 1 and 2, note that the present modeling is not meant to reproduce actual rates in E galaxies, but to illustrate general trends.

Another possible way in which age might affect the characteristics of the WD progenitors of SN Ia is *white dwarf cooling*. If the exploding WDs in E galaxies all formed $\sim 10^{10}$ *yr* ago, most of their interiors should have crystallized. That allows a range of ignition densities within a factor $\simeq$ 2–3 (Hernanz et al. 1988). The changes in the explosion characteristics induced by changes in the explosive ignition density have been explored by Bravo et al. (1993), for two different burning propagation mechanisms (deflagrations and nonpulsating delayed detonations). The explosion energies remain almost constant. Changes in bolometric magnitudes would be in the range 0.02–0.40 mag. Variations in photospheric velocities would have an amplitude of $\sim 800 - 1,400$ *km s*$^{-1}$, but they would *correlate inversely* with luminosity: the brighter events would have the lower photospheric velocities around maximum light. Equally, despite variations in the mass of $^{56}$Ni in pulsating delayed detonation models, these models also give nearly constant explosion energies (see Wheeler et al. 1994). The relationship between the density at which the detonation starts (that controls the $^{56}$Ni mass) in these models with the properties of the binary system remains unexplored. All those considerations would apply not only to the DD scenario but also to all scenarios involving the explosion of a Chandrasekhar–mass WD.

We have explored, in the preceding, the effects leading to possible correlations of luminosity with expansion velocity and galaxy type in SNe Ia. The most relevant factor appears to be the age difference among the dominant stellar populations. Different effects are expected, depending on the adopted scenario for the evolution of the SN Ia progenitors and on whether the exploding WD has reached the Chandrasekhar mass or not. *The SD scenario in which WDs with masses below the Chandrasekhar mass explode in symbiotic systems appears as the most promising one to explain the observed trends: decreasing SN Ia rate, fainter outbursts, and smaller expansion velocities when going from late towards early–type galaxies.* The DD scenario alone could maybe explain (allowing for uncertainties in the $\alpha$ parameter discussed above) the lower SN Ia rate in early–type galaxies and maybe also, via higher ignition densities due to cooler WD progenitors, the lower luminosities of the events (if the differences were not much larger than $\sim$ 0.1–0.3 mag). It is not clear, however, which physical properties of the DD scenario could produce the observed correlations. In explosion models involving a Chandrasekhar–mass WD, the spread in the decline of the light curve and in maximum brightness results from differences in the propagation of the thermonuclear flame across the WD. Any possible physical relationship between the dynamics of burning and the parent population of the exploding WD remains unknown. Another possible way to explain the observed correlations might be to invoke a mixture of two types of progenitors: DD (with Chandrasekhar masses at explosion) and SD (with masses below the Chandrasekhar mass). The former would be brighter in average and more energetic and their rate would decrease faster with age than that of the latter, in such a way that the low–luminosity SD would dominate in E galaxies. More complex modeling (work in progress) will explore this issue. For all scenarios, the SN Ia rate in E galaxies should increase with redshift $z$, reaching a maximum at a value of $z$ which depends on the adopted scenario. This can be seen in Figure 1: in the symbiotic scenario the maximum rate is reached $\sim 10^9$ yr after the end of star formation, decreases more slowly than in the other scenarios and reaches a second peak at about $5 \times 10^9$ yr. The recent work by Yungelson et al. (1995) agrees with these results in the time evolution and in the location of the peaks. Their absolute scale takes into account the efficiency of accretion (it is set to unity in the present work). Since most systems explode at t $\lesssim 6 \times 10^9$ yr, they label them as relatively "young", but they are rather "old" when compared with the other scenarios. In the CO + He star scenario the rate would be peaking around $2 \times 10^8$ yr and in the CO +CO scenario (depending on the common envelope parameter) at about $10^8$ yr. If we adopt a specific galaxy formation model we can estimate at which redshift these maxima should be expected (Figure 2). For the standard CDM model (with $h = 0.5$), in which galaxies of about $10^{12}$ $M_\odot$ form at redshift z~5, the maxima in the SNe Ia rate for the symbiotic scenario take place at z~2 (absolute maximum) and z~0.6 (lower peak), whereas in the CO + He scenario and the CO + CO scenario these peaks are expected earlier: at z~4 and z~4.4 respectively. These results do not depend on the assumed IMF, as can be seen in Figure 2. In the CO + He star scenario and in the CO + CO scenario the decrease after peak is steeper than for symbiotic systems. The different shapes of the evolution of SNe Ia rates with redshift for the different scenarios might be testable through comparisons up to moderate redshifts z~0.6 (the statistics, however, is only good nowadays up to z~0.1). *If explosion of WDs with masses below the Chandrasekhar mass are involved, the average brightness of the SN Ia should increase with z.* This is due to the fact that the average mass of the exploding WD monotonously decreases with time, as can be seen in the second panel of Figure 1. That should also be expected for a combination of the two scenarios (symbiotic SD and DD): not only the contribution from symbiotics for increasing z would correspond to more massive WDs, but also the ratio of the rates SD/DD would decrease (they fall more steeply with time for the DD scenario), that increasing the relative contribution of Chandrasekhar–mass WDs. The trend of brighter events corresponding to younger populations, if confirmed along the Hubble sequence (E–S0–Sa–Sb–Sc) would give additional support to the suggestion that a sub–Chandrasekhar explosion mechanism is at work. The observed variation of velocity of the ejecta along the Hubble sequence (Branch & van den Bergh 1993) is quite significant, already, on this respect.

The observed spread in magnitudes and rate of decline of the light curves of SNe Ia has suggested a modified use of SNe Ia as "standard candles", through the correlation of the maximum brightness and rate of decline of the light curve (Phillips 1993; Riess, Press, & Kirshner 1994). The question has been raised of whether a second parameter related to the population would be involved in the maximum brightness versus rate of decline relationship and would hinder its use in distance determinations. If only the single degenerate scenario were at work (through sub–Chandrasekhar explosions), the brightness of SNe Ia should be a single parameter class, well described by the rate of decline of the light curve or by the expansion velocity of the ejecta. The environmental effects are pure population age effects acting through the mass distribution of the WD progenitors, and there should not be any further major dependence from environment. If both the single degenerate scenario and the double degenerate scenario do contribute to Type Ia explosions, then such relationship should be blurred by the mixing of two different families of behavior. Ellipticals should contain a single family of explosions (sub–Chandrasekhar) and thus display a sharper kinematic–brightness relationship along redshift. Therefore, despite enviromental effects, SNe Ia brightness at maximum would keep its use as a fruitful distance indicator, although the path through it towards $H_0$ and $q_0$ becomes somewhat more laborious.

We thank the referee, David Branch, for useful suggestions to improve the manuscript.

# REFERENCES

Branch, D., & van den Bergh, S. 1993, AJ, 105, 2231

Bravo, E., et al. 1993, A&A, 269, 187

Della Valle, M., & Livio, M. 1994, ApJ, 423, L31

Duquennoy, A., & Mayor, M. 1991, A&A, 284, 485

Filippenko, A.V. 1989, PASP, 101, 588

Hamuy, M., et al. 1995, AJ, 109, 1

Hernanz, M., Isern, J., Canal, R., Labay, J., & Mochkovitch, R. 1988, ApJ, 324, 331

Iben, I.,Jr., & Renzini, A. 1983, ARA&A, 21, 271

Iben, I.,Jr., & Tutukov, A.V. 1984, ApJS, 54, 335

——————————. 1991, ApJ, 370, 615

Kenyon, S.J., Livio, M., Mikolajewska, J., & Tout, C.A. 1993, ApJ, 407, L81

Landau, L., & Lifshitz, E. 1975, The Classical Theory of Fields (London: Pergamon)

Livio, M., & Soker, N. 1988, ApJ, 329, 764

Livne, E., & Glasner, A.S. 1991, ApJ, 370, 272

Maza, J., Hamuy, M., Phillips, M.M., Suntzeff, N.B., & Avilés, R. 1994, ApJ, 424, L107

Munari, U., & Renzini, A. 1992, ApJ, 397, L87

Phillips, M.M. 1993, ApJ, 413, L105

Popova, E.I., Tutukov, A.V., & Yungelson, L.R. 1982, Ap&SS, 88, 155

Riess, A., Press, W., & Kirshner, R.P. 1995, ApJ, in press

Roberts, M.S., & Haynes, M.P. 1994, ARA&A, 32, 115

Ruiz–Lapuente, P., et al. 1993, Nature, 365, 728

Salpeter, E.E. 1955, ApJ, 121, 161

Scalo, J.M. 1986, Fundamentals of Cosmic Physics, 11, 1

Schaller, G., Schaerer, D., Meynet, G., & Maeder, A. 1992, A & AS, 96, 269

Taam, R.E., & Bodenheimer, P. 1989, ApJ, 337, 849

Tutukov, A.V., & Yungelson, L.R. 1994, MNRAS, 268, 871

van den Bergh, S., & Pazder, J. 1992, ApJ, 390, 34

Wheeler, J.C., Harkness, R.P., Khokhlov, A.M., & Höflich, P.A. 1994, preprint

Woosley, S.E., & Weaver, T.A. 1994, ApJ, 423, 371

Yungelson, L.R., Livio, M., Tutukov, A.V., & Saffer, R.A. 1994, ApJ, 420, 336

Yungelson, L.R., Livio, M., Tutukov, A.V., & Kenyon, S.J. 1995, preprint

## Figure Captions

*Figure 1.* From the top to bottom panel: a) Evolution of the SN Ia rate after an outburst of star formation, in the SD (symbiotic) scenario. b) Time evolution of the average mass of the exploding WD (in $M_\odot$), for the symbiotic scenario. c) Time evolution of the average expansion velocity (in $10^4$ km s$^{-1}$) of SNe Ia, in the symbiotic scenario. d) Same as a), for the SD (CO WD + He star) scenario. Cases corresponding to the common–envelope parameter $\alpha = 0.5$ (dashed line), 1 (solid line), and 2 (dot-dashed line) are shown. e) Same as d), for the DD (CO WD + CO WD) scenario.

*Figure 2.* SN Ia rates for E galaxies as a function of redshift. Formation of the E galaxies is assumed to happen at z ∼5 (see text). In the top and central panels: the solid line corresponds to the Scalo (1986) IMF and the $A_0$ distribution of Duquennoy & Mayor (1991). The dot–dashed line, to the IMF of Salpeter (1955) and the $A_0$ distribution of Popova, Tutukov & Yungelson (1982). Rates are shown for the CO +CO WD scenario (common envelope parameter $\alpha$= 1) and the CO WD + He star scenario (top panel), and for the symbiotic scenario (central panel). The bottom panel shows the predictions up to redshifts accessible to observations. The rates are given in the galaxy frame. For the symbiotic scenario, both the case of an E0 galaxy (solid line) with a burst of star formation lasting $10^9$ yr and the case of an instantaneous burst (dot–dashed line) are shown. A peak at z ∼ 0.5–0.7 is found for this scenario. The DD (dashed line) and CO + He star (dotted line) scenarios show a steady decrease. The absolute scale has been shifted to allow comparison.

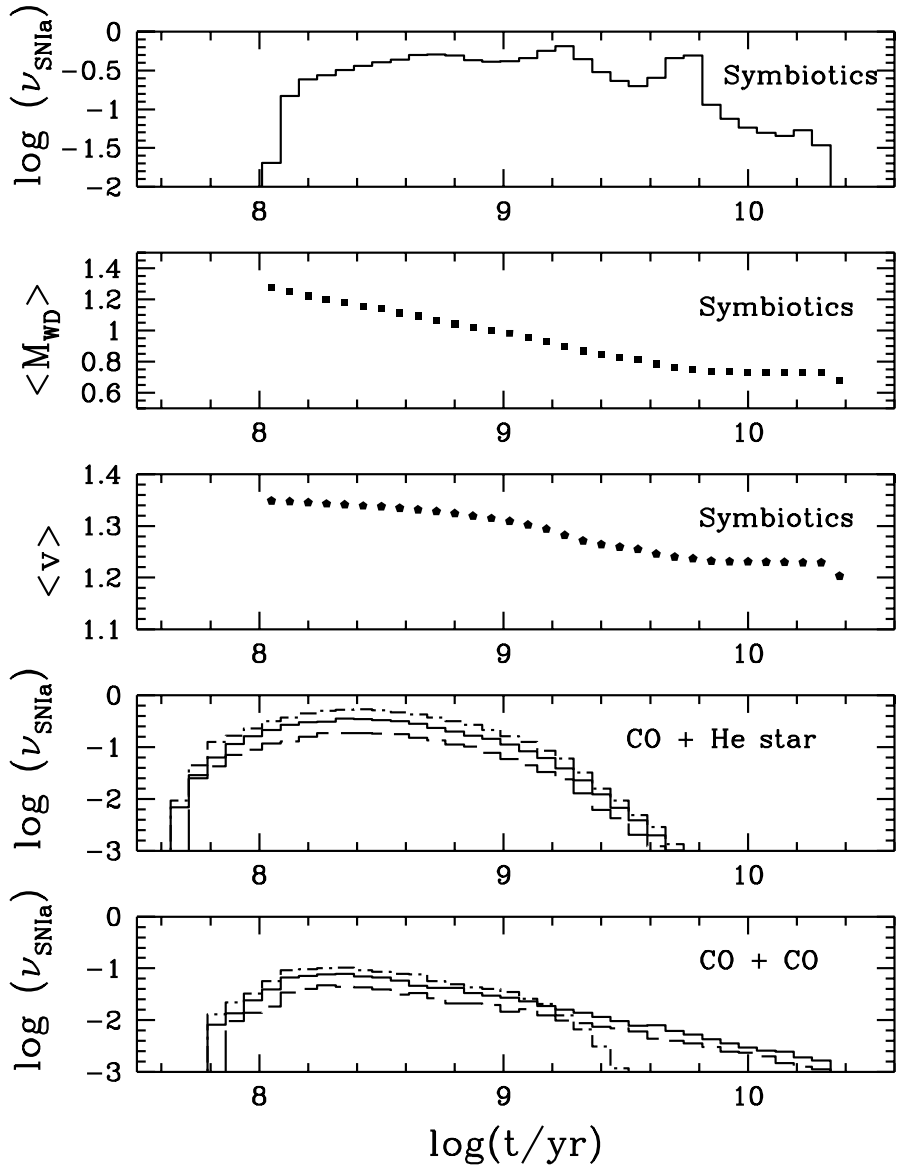

Fig. 1.—

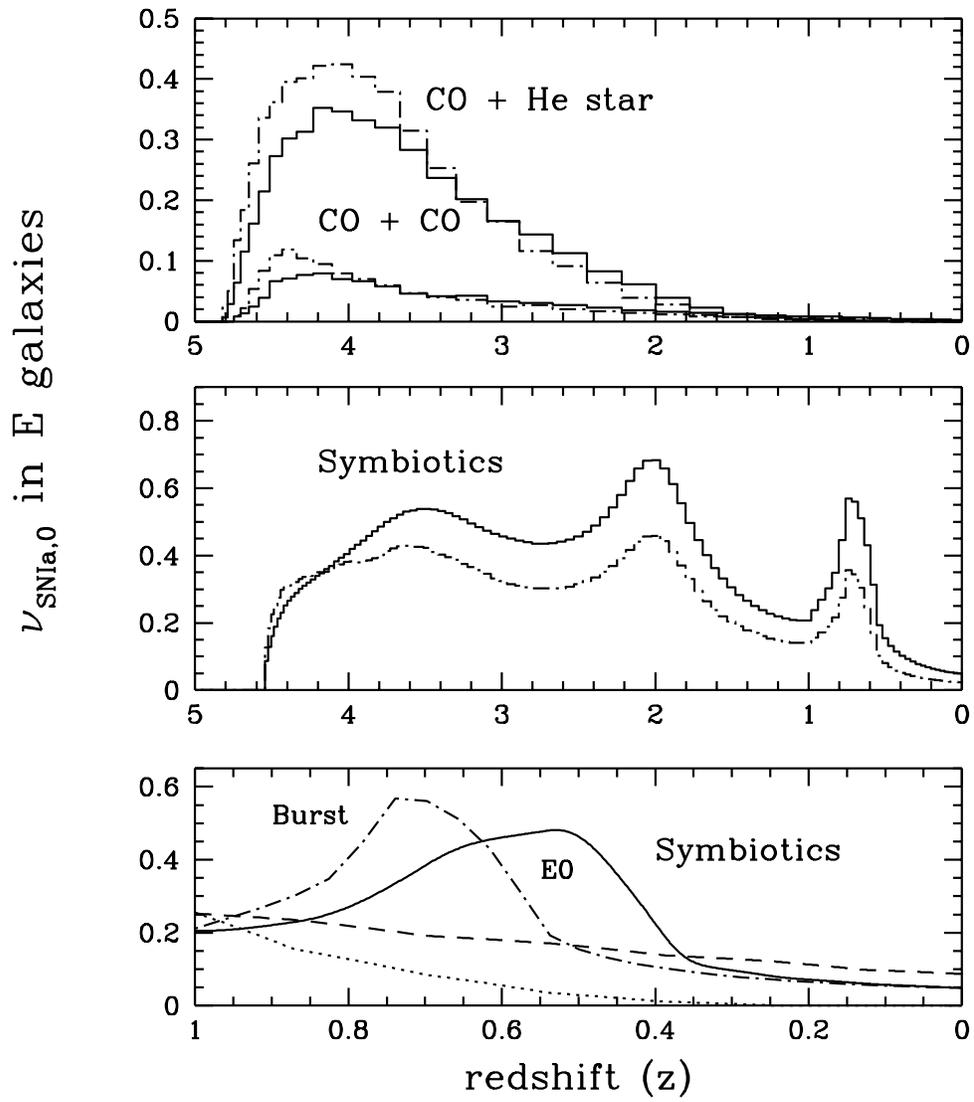

Fig. 2.—